\documentclass[12pt]{article}
\usepackage{amsmath}
\usepackage{amsfonts}
\newcommand{\be}{\begin{equation}}
\newcommand{\ee}{\end{equation}}
\newcommand{\ba}{\begin{eqnarray}}
\newcommand{\ea}{\end{eqnarray}}
\begin{document}

\title{Hidden symmetries in a gauge covariant approach, Hamiltonian 
reduction and oxidation}

\author{Mihai Visinescu \thanks{E-mail:~~~mvisin@theory.nipne.ro}\\
{\small \it Department of Theoretical Physics,}\\
{\small \it Horia Hulubei National Institute for Physics and Nuclear 
Engineering,}\\
{\small \it P.O.B. MG-6, 077125 Magurele, Romania}}
\date{}

\maketitle

\begin{abstract}
Hidden symmetries in a covariant Hamiltonian formulation are 
investigated involving gauge covariant equations of motion.
The special role of the St\"ackel-Killing tensors is pointed out. 
A reduction procedure is used to reduce the original phase space to 
another one in which the symmetries are divided out. The reverse of 
the reduction procedure is done by stages performing the unfolding of 
the gauge transformation followed by the Eisenhart lift in connection 
with scalar potentials.
\end{abstract}

\section{Introduction}

The motion of a free point particle in a (pseudo)-Riemannian 
space is determined by the kinetic energy and the 
trajectories are geodesics corresponding to the metric tensor of 
the configuration space. More general than the free mechanical system, 
in the case of a conservative holonomic dynamical system whose kinetic 
energy is modified by the addition of a potential, the trajectories 
are not geodesics of the metric tensor.

In many cases it is preferably to represent the trajectories as 
geodesics of a related metric. For example the Jacobi metric \cite{AM}, 
conformally related to the original one,
is obtained rescaling the potential and the total energy. The drawback 
of the conformally related metric is that its geodesics describe the 
trajectories of a fixed energy.

An attractive alternative is represented by the Eisenhart lift
or oxidation \cite{LPE} of a dynamical system which 
permits to put into correspondence the trajectories of a mechanical 
system with the geodesics of a configuration space extended in 
dimension.

It is well known that a geodesic system has a first integral linear 
in the momenta if the metric admits a Killing vector corresponding to 
an infinitesimal isometry. In some cases there exist additional 
nontrivial first integrals quadratic (or more general polynomial) in 
the momenta provided that the configuration space admits 
St\"ackel-Killing (SK)
tensors. Superintegrable systems have been thoroughly studied in 
connection with hidden symmetries and separability of the 
Hamilton-Jacobi equations and the corresponding ones in the quantum 
theory. From this point of view the superintegrable systems, 
formulated as geodesic systems, offer us examples of manifolds with 
nontrivial Killing tensors.

In this paper we analyze the hidden symmetries of a 
dynamical system in the presence of external gauge fields in a 
covariant approach \cite{vH,JPN,MV1}. This approach proves to be more 
convenient in the study of the conserved quantities involving gauge 
covariant equations of motion.

In the case of a symplectic manifold on which a group of symmetries 
acts symplectically, it is possible to reduce the original phase space 
to another symplectic manifold in which the symmetries are divided out. 
Such a situation arises when one has a particle moving in an 
electromagnetic field \cite{MW}.

On the other hand the reverse of the reduction procedure can 
be used to investigate complicated systems. It is possible to use a sort 
of unfolding of the initial dynamics by imbedding it in a larger one 
which is easier to integrate \cite{MSS}. Sometimes the equations of 
motion in a higher dimensional space are quite transparent, e. g. 
geodesic motions, but the equations of motion of the reduced system 
appear more complicated \cite{KKS}.

As an illustration of the reduction of a symplectic manifold with 
symmetries and the opposite procedure of oxidation of a dynamical 
system we shall consider the principal bundle
$\pi:\mathbb{R}^4 - \{0\} \rightarrow \mathbb{R}^3 - \{0\}$ 
with structure group $U(1)$. The Hamiltonian function on
the cotangent bundle $T^\star({\mathbb{R}}^4 -\{0\})$ is invariant 
under the $U(1)$ action and the reduced Hamiltonian system proves to 
describe the three-dimensional Kepler problem in the presence of a 
centrifugal potential and Dirac's monopole field. Moreover this 
reduction procedure is also relevant for many other problems like the 
geodesic flows of the generalized Taub-NUT metric, conformal Kepler 
system, MIC-Kepler system, etc.

Concerning the unfolding of the reduced Hamiltonian system we shall 
perform it by stages. In a first stage of unfolding we use an opposite 
procedure to the reduction by an $U(1) \simeq S^1$ action to a 
four-dimensional generalized Kepler problem. Finally we resort to 
the method introduced by Eisenhart who added one or two extra 
dimensions to configuration space to 
represent trajectories by geodesics. The kinetic energy metric and 
scalar potential are involved in the construction of the metric of the 
extended configuration space in such a way that geodesics on the 
extended space project to trajectories of the initial configuration 
space.

The plan of the paper is as follows. Section 2 is concerned with the 
covariant formulation of the dynamics of particles in the presence of 
external gauge fields and scalar potentials. In Section 3 we make a 
brief review of the Hamiltonian reduction of symplectic manifolds with 
symmetries pointing out the Hamiltonian systems defined on the 
cotangent bundle $T^\star({\mathbb{R}}^4 -\{0\})$ with standard 
symplectic form. Next Section is 
devoted to the  reverse of the reduction procedure associated with an 
$S^1$ action to a four dimensional generalized Kepler problem. In 
Section 5 we discuss the Eisenhart procedure for the oxidation limiting 
ourselves to dynamical systems and scalar potentials which do not 
involve time. Conclusions and open problems are discussed in the last 
Section.

\section{Symmetries and conserved quantities}

The geodesic flow for an $n$-dimensional manifold $M$ equipped with 
a (pseudo)-Riem\-ma\-ni\-an metric $\mathbf{g}$ is generated by the 
quadratic Hamiltonian
\be\label{H1}
H = \frac{1}{2} g^{ij} p_i p_j \,.
\ee
In terms of the phase-space variables $(x^i, p_i)$ the canonical 
symplectic structure $\omega$ of $T^\star M$ is $\omega = 
dp_i\wedge dx^i$ and the corresponding Poisson bracket of two 
observables $P, Q$ is 
\be\label{Pb}
\{P,Q\} = \frac{\partial P}{\partial x^i} \frac{\partial Q}
{\partial p_i} -\frac{\partial P}{\partial p_i} \frac{\partial Q}
{\partial x^i} \,.
\ee

Let us consider a conserved quantity of motion expanded as a power 
series in momenta:
\be\label{cq}
K =  \sum_{i=0}^s K^{(i)}=
 K_0  + \sum^{s}_{k=1}\frac{1}{k!} K^{i_1 \cdots i_k} (x) 
p_{i_1} \cdots p_{i_k}\,.
\ee
with $K^{(i)}$ homogeneous polynomial of degree $i$ in momenta.
It has vanishing Poisson bracket with the Hamiltonian, $\{K,H\} = 0$, 
which implies
\be\label{SKT}
K^{(i_1 \cdots i_k;i)} =0\,,
\ee
where a semicolon denotes the covariant differentiation corresponding 
to the Levi-Civita connection $\nabla$  and round brackets indicate full 
symmetrization over the indices enclosed. A symmetric tensor
$K^{i_1 \cdots i_k}$ satisfying \eqref{SKT} is called a 
SK tensor of rank $k$. The SK tensors represent a 
generalization of the Killing vectors and are responsible for the hidden 
symmetries of the motions, connected with conserved quantities of the 
form \eqref{cq} polynomials in momenta. 

The traditional means to deal with the coupling to a gauge field 
$F_{ij}$ expressed (locally) in terms of the potential $1$-form 
$A_{i}$
\be\label{FdA}
F=dA \,,
\ee
is to replace the Hamiltonian by
\be\label{H2}
H = \frac{1}{2} g^{ij} (p_i - A_i) (p_j - A_j)  + V(x) \,,
\ee
work with the Poisson bracket \eqref{Pb} and consider the 
polynomials \eqref{cq} in the variables $(p_i - A_i)$ for $i= 1, 
\cdots ,n$ \cite{DV}. For completeness, in \eqref{H2} we included a 
scalar potential $V(x)$.

The disadvantage of this approach is that the canonical momenta $p_i$ 
and implicitly the Hamilton equations of motion are not manifestly 
gauge covariant. This inconvenience can be removed using van Holten's 
receipt \cite{vH} by introducing the gauge invariant momenta:
\be\label{Pi}
\Pi_i = p_i - A_i \,.
\ee
The Hamiltonian \eqref{H2} becomes
\be\label{H3}
H = \frac{1}{2} g^{ij} \Pi_i \Pi_j  + V(x) \,,
\ee
and the equations of motion are derived using the modified Poisson 
bracket
\cite{vH,DV,JMS}
\be\label{covPB}
\{P,Q\} = \frac{\partial P}{\partial x^i} \frac{\partial Q}{\partial 
\Pi_i} -\frac{\partial P}{\partial \Pi_i} \frac{\partial Q}{\partial x^i} 
+ q F_{ij}\frac{\partial P}{\partial \Pi_i} 
\frac{\partial Q}{\partial \Pi_j} \,.
\ee

Searching for conserved quantities \eqref{cq} expanded rather into 
powers of the gauge invariant momenta $\Pi_i$, the vanishing of the 
Poisson bracket $\{K,H\}$ yields a series of constraints in the form of 
a system of coupled differential equations \cite{MV1}. Only the 
equation for the leading order term $K^{i_1 \cdots i_s}$ defines a SK 
tensor of rank $s$. The rest of equations mixes the derivatives of the 
terms $K^{i_1 \cdots i_k} \,, (k<s)$ and potential $V$ with the 
gauge field strength $F_{ij}$.

Several applications using van Holten's covariant framework \cite{vH}
are given in \cite{JPN,MV1,HN,IKI,MV3}.

\section{Hamiltonian reduction}

It is simplest to work with the Hamiltonian formulation in order to see 
how the reduction and the oxidation of a dynamical system affect 
constants of motion. 

The general setting for reduction of symplectic manifolds with 
symmetries is presented in \cite{AM,MW}. 
Here we confine ourselves to the $U(1)$ reduction 
of a Hamiltonian system defined on the cotangent bundle 
$T^\star({\mathbb{R}}^4 -\{0\})$ with standard symplectic form. 
The reduced phase space is not symplectmorphic to the cotangent bundle
$T^\star({\mathbb{R}}^3 -\{0\})$ with standard symplectic form.
It proves that the reduced symplectic form on 
$T^\star({\mathbb{R}}^3 -\{0\})$ contains a two-form describing 
Dirac's monopole field beside the standard symplectic form.

Let us start to consider the principal fiber bundle 
$\pi:\mathbb{R}^4 - \{0\} \rightarrow \mathbb{R}^3 - \{0\}$ 
with structure group $U(1)$ whose action is given by \cite{IU}
\be
x \mapsto T(t) x \,,\quad x\in \mathbb{R}^4 \,,
\quad t\in \mathbb{R} \,,
\ee
where
\be
T(t) = \left(\begin{array}{cc} R(t) & 0\\ 0 & R(t) \end{array}  
\right )\,, \quad
R(t) = \left(\begin{array}{cc} \cos \frac{t}{2} & -\sin \frac{t}{2}\\ 
\sin \frac{t}{2} & \cos \frac{t}{2}  \end{array} \right )\,.
\ee

The $U(1)$ action is lifted to a symplectic action on 
$T^\star( \mathbb{R}^4 - \{0\})$
\be\label{symplact}
(x,y) \rightarrow (T(t) x, T(t) y) ,\quad (x,y) \in( \mathbb{R}^4 - 
\{0\}) \times \mathbb{R}^4 \,.
\ee

Let $ \Psi : T^\star( \mathbb{R}^4 - \{0\})\rightarrow \mathbb{R}$
be the moment map associated with the $U(1)$  action \eqref{symplact}
\be 
\Psi (x,y) = \frac{1}{2} (- x_2 y_1 + x_1 y_2 -x_4 y_3 + x_3 y_4)\,.
\ee

The reduced phase-space $P_\mu$ is defined through 
\be
\pi_\mu : \Psi^{-1} (\mu) \rightarrow P_\mu := \Psi^{-1}(\mu) /U(1)\,,
\ee
which is diffeomorphic  with $T^\star( \mathbb{R}^3 - \{0\})
\cong ( \mathbb{R}^3 - \{0\}) \times \mathbb{R}^3$.

The coordinates $(q_k, p_k) \in (\mathbb{R}^3 - \{ 0 \}) \times 
\mathbb{R}^3$ are given by the Kustaan\-hei\-mo-Stiefel 
transformation
\be
\left(\begin{array}{c}q_1\\ q_2\\ q_3\\0\end{array}\right) =
\left( \begin{array}{rrrr} x_3 & x_4 & x_1 & x_2 \\
-x_4 & x_3 & x_2 & -x_1 \\
x_1 & x_2 & -x_3 & -x_4 \\
-x_2 & x_1 & -x_4 & x_3 \end{array}\right)
\left(\begin{array}{c}x_1\\ x_2\\ x_3\\x_4 \end{array}\right)\,,
\ee
\be
\left(\begin{array}{c}p_1\\ p_2\\ p_3\\\Psi /r
\end{array}\right) = \frac{1}{2r}
\left( \begin{array}{rrrr} x_3 & x_4 & x_1 & x_2 \\
-x_4 & x_3 & x_2 & -x_1 \\
x_1 & x_2 & -x_3 & -x_4 \\
-x_2 & x_1 & -x_4 & x_3 \end{array}\right)
\left(\begin{array}{c}y_1\\ y_2\\ y_3\\y_4 \end{array}\right)\,.
\ee
where $r = \sum^{4}_{1}x^2_j =\sqrt{ \sum^{3}_{1} q^2_k} $.

The phase-space $T^\star( \mathbb{R}^4 - \{0\})$ is equipped with the 
standard symplectic form
\be
d\Theta = \sum^{4}_{1} d y_j \wedge dx_j \,,\quad 
\Theta  = \sum^{4}_{1} y_j \wedge dx_j\,.
\ee

Let $\iota_\mu : \Psi^{-1}(\mu) \rightarrow T^\star( \mathbb{R}^4 - 
\{0\})$ be the inclusion map. The reduced symplectic form $\omega_\mu$
is determined on $P_\mu$ by
\be
\pi^\star_\mu \omega_\mu = \iota^\star_\mu d\Theta\,,
\ee
namely
\be\label{sym}
\omega_\mu = \sum^{3}_{k=1} dp_k \wedge dq_k - \frac{\mu}{r^3}
(q_1\, dq_2 \wedge dq_3 + q_2\, dq_3 \wedge dq_1 + q_3\, dq_1 \wedge 
dq_2)\,.
\ee  
$\omega_\mu$ consists of the standard symplectic form on 
$T^\star( \mathbb{R}^3 - \{0\})$ and in addition a term corresponding 
to the  Dirac's monopole field
\be\label{Bmonopole}
\vec{B} = - \mu \frac{\vec{q}}{r^3}\,,
\ee
of strength $-\mu$.

The reduced Hamiltonian is determined by
\be\label{redH}
H\circ \iota_\mu = H_\mu \circ \pi_\mu\,.
\ee

For the purpose of the present 
paper, we shall be concerned with the reduction of the dynamical system 
associated with the geodesic flows of the generalized Taub-NUT metric 
on ${\mathbb{R}}^4 -\{0\}$. This metric is  relevant for (conformal) 
Coulomb problem \cite{IU}, MIC-Zwanziger system \cite{DZ,MIC}, 
Euclidean Taub-NUT \cite{SWH,NSM, GR} and its extensions \cite{IK1,IK2}, 
etc. The generalized Taub-NUT metric is
\be
ds^2_4 = f(r) (dr^2 + r^2 (d\theta^2 + \sin^2\theta d\phi^2)) + g(r) 
(d\psi + \cos \theta d \phi)^2\,,
\ee
where the curvilinear coordinates $(r,\theta,\phi,\psi)$ are
\ba
x_1 = \sqrt{r} \cos\frac{\theta}{2} \cos \frac {\psi + \phi}{2} \,, &&
x_2 = \sqrt{r} \cos\frac{\theta}{2} \sin  \frac {\psi + 
\phi}{2}\,,\nonumber\\
x_3 = \sqrt{r} \sin\frac{\theta}{2} \cos \frac {\psi - \phi}{2} \,,&&
x_4 = \sqrt{r} \sin\frac{\theta}{2} \sin \frac {\psi - \phi}{2}\,.
\ea

In what follows we consider the  Hamiltonian on the cotangent bundle 
$T^\star( \mathbb{R}^4 - \{0\})$ 
\be\label{H4}
H =\frac{1}{2 f(r)} p^2_r + \frac{1}{2 r^2 f(r)}p^2_\theta 
+ \frac{(p_\phi - p_\psi \cos\theta)^2}{2 r^2 f(r) \sin^2\theta}
+ \frac{p_\psi^2}{2 g(r)} + V(r) \,,
\ee
where, to make things more specific, we consider a potential $V(r)$ 
function of the radial coordinate $r$.

The Hamiltonian function is invariant under the $U(1)$ action with the 
infinitesimal generator $\frac{\partial}{\partial \psi}$ so that the 
conserved momentum is
\be
\mu = p_\psi  = \Theta (\frac{\partial}{\partial \psi}) \,,
\ee
where the canonical one-form $\Theta$ could be expressed in curvilinear 
coordinates
\be\label{Theta}
\Theta = p_r dr + p_\theta d\theta + p_\phi d\phi + p_\psi d\psi\,,
\ee
on the cotangent bundle $T^\star( \mathbb{R}^4 - \{0\})$.

The reduced Hamiltonian \eqref{redH} has the form
\be\label{Hmu}
H_\mu = \frac{1}{2f(r)} \sum^{3}_{k=1} p_k^2 + \frac{\mu^2}{2g(r)} + 
V(r)\,.
\ee

Now the search of conserved quantities of motion in the $3$-dimensional 
curved space in the presence of the potential $V(r)$ plus the 
contribution of the monopole field proceeds in standard way. 
First of all we remark that the reduced Hamiltonian is still spherical 
symmetric and one can easily show that the angular momentum vector 
\be
\vec{J} = \vec{q} \times \vec{p} + \frac{\mu}{r} \vec{q}\,,
\ee
is conserved.

In some cases the system admits additional constants of motion 
polynomial in momenta. Here are some notable cases:

1) For 
$$f(r) =1\,,\quad g(r) = r^2\,,\quad V(r) = -\frac{\kappa}{r}$$ 
we recognize the MIC-Kepler problem with the Runge-Lenz type 
conserved vector
\be
\vec{A} = \vec{p} \times \vec{J} - \kappa \frac{\vec{q}}{r}\,.
\ee

1a) Moreover, for $\mu=0 \,, H_\mu$ becomes the Hamiltonian for the 
Coulomb - Kepler problem.

2) For 
$$f(r) = \frac{a+b r}{r}\,,\quad g(r) = \frac{a r + b r^2} 
{1 + c r + d r^2}\,,\quad V(r) = 0\,,$$
with $a, b, c, d$ real constants 
we recover the extended Taub-NUT space which still admit a Runge-Lenz 
type vector

\be
\vec{A} = \vec{p} \times \vec{J} - (a E - \frac{1}{2} c \mu^2) 
\frac{\vec{q}}{r}\,,
\ee
where $E$ is the conserved energy.

2a) In the particular case, if the constants $a, b, c, d$ are subject 
to the constraints
\be
c=\frac{2b}{a}\,,\quad d = \frac{b^2}{a^2}\,,
\ee
the extended metric coincides, up to a constant factor, with the 
original Taub-NUT metric.

Other examples could be found in \cite{JPN,IK1,GW}.

\section{Unfolding}

It is interesting to analyze the reverse of the reduction procedure 
which can be used to investigate difficult problems \cite{MSS}.
For example the 
equations of motion for the dynamical system \eqref{sym}, \eqref{Hmu}
look quite complicated. Using a sort of {\it unfolding} of the 
$3$-dimensional dynamics imbedding it in a higher dimensional space 
the conserved quantities are related to the symmetries of 
this manifold.

To exemplify let us start with the reduced Hamiltonian \eqref{Hmu} 
written in curvilinear coordinates
\be\label{Hmusph}
H_\mu =\frac{1}{2 f(r)} \left [p_r^2 + \frac{1}{r^2}\left( p_\theta^2 + 
\frac {(p_\phi - \mu \cos\theta)^2}{\sin^2\theta}\right)\right] + 
\frac{\mu^2}{2 g(r)} + V(r) \,,
\ee
on the $3$-dimensional space with the metric
\be
ds^2_3 = f(r) (d r^2 + r^2 (d\theta^2 + \sin^2\theta d\phi^2))\,,
\ee 
and  the canonical symplectic form
\be
d\Theta_\mu = dp_r \wedge dr + dp_\theta \wedge d\theta + dp_\phi \wedge 
d\phi\,,
\ee
$\mu$ being the strength of the Dirac's monopole field.

In the specific case of the Dirac's monopole field \eqref{Bmonopole} 
the gauge invariant momenta \eqref{Pi} in spherical coordinates are
\be
\Pi_r = p_r \,, \quad \Pi_\theta = p_\theta \,, \quad \Pi_\phi= p_\phi 
- \mu \cos \theta\,.
\ee
The reduced Hamiltonian \eqref{Hmu} has the form
\be\label{Hmucov}
H_\mu =\frac{1}{2 f(r)} \left [\Pi_r^2 + \frac{1}{r^2}\left( 
\Pi_\theta^2 + \frac {\Pi_\phi^2}{\sin^2\theta}\right)\right] + 
\frac{\mu^2}{2 g(r)} + V(r) \,,
\ee
and
\be\label{symcov}
d\Theta_\mu = d\Pi_r \wedge dr + d\Pi_\theta \wedge d\theta + 
d\Pi_\phi \wedge d\phi - \mu \sin\theta d\theta \wedge d\phi \,,
\ee
in  agreement with \eqref{sym} and  Poisson bracket \eqref{covPB}.

At each point of $T^\star( \mathbb{R}^3 - \{0\})$ we define the fiber 
$S^1$, the group space of the gauge group $U(1)$.
On the fiber we consider the motion whose equation is
\be
\frac{d\psi}{dt} = \frac{\mu}{g(r)} - \frac{\cos\theta}{r^2 f(r)
\sin^2\theta}(p_\phi -\mu\cos\theta)\,.
\ee

The metric on $\mathbb{R}^4$ defines horizontal spaces orthogonal to 
the orbits of the circle - this is a connection on the principal 
bundle \cite{NH}. Using the above trivialization, we have the 
coordinates $(r, \theta,\phi, \psi)$ with the horizontal spaces 
annihilated by the connection
\be
d\psi + \cos\theta d\phi\,.
\ee

The metric on $\mathbb{R}^4$, which admits a circle action leaving 
invariant the symplectic form \eqref{symcov}, can be written in the form
\ba
ds^2_4 & = & f(r) (d r^2 + r^2 (d\theta^2 + \sin^2\theta d\phi^2)) 
+ h(r) (d\psi + \cos\theta d\phi)^2\nonumber\\
& = & \sum^4_{i,j = 1} g_{ij}dq^i dq^j\,.
\ea 
The natural symplectic form on $T^\star( \mathbb{R}^4 - \{0\})$ is
\eqref{Theta}
\be
d\Theta = d\Theta_\mu + dp_\psi \wedge d\psi\,.
\ee

Considering the geodesic flow of $ds^2_4$ and taking into account that 
$\psi$ is a cycle variable
\be
p_\psi = h(r) (\dot\psi + \cos\theta \dot\phi)\,,
\ee
is a conserved quantity. To make contact with the Hamiltonian dynamics 
on $T^\star( \mathbb{R}^3 - \{0\})$ we must  identify
\be\label{hg}
h(r)=g(r)\,.
\ee
Otherwise the resulting Hamiltonian dynamics projected onto
 $T^\star( \mathbb{R}^3 - \{0\})$ is that from the Hamiltonian $H_\mu$ 
choosing \eqref{hg}.

The reverse of the reduction proves to be useful since the equations 
of motion for the Hamiltonian
\be\label{Hunfold}
H = \frac{1}{2} g^{ij} p_i p_j +V\,,
\ee
are quite simple and transparent, but the equations of the quotient 
system \eqref{Hmusph} appear more complicated \cite{IK2}. The 
corresponding differential equations of the trajectories are
\be\label{geod}
g_{ij} \ddot{q}^j + [jk,i] \dot{q}^j \dot{q}^k  + 
\frac{\partial V}{\partial q^i} = 0 \,,
\ee
where $[jk,i]$ is the Christoffel symbol. These equations admit the 
first integral of motion
\be\label{TVE}
\frac{1}{2} g_{ij} \dot{q}^i \dot{q}^j + V = T + V = E\,,
\ee
where $E$ is the conserved energy.

\section{Eisenhart lift}

In many concrete problems, after the unfolding of the gauge symmetry, 
one ends up with a dynamical system on an extended phase space and an 
Hamiltonian \eqref{Hunfold} with a "residual" scalar potential.

In the final stage of the oxidation of the dynamical system described 
by the Hamiltonian \eqref{Hunfold} we shall apply the Eisenhart's lift 
\cite{LPE} (see also \cite{MS,IMB}). In the general case of the 
Eisenhart's lift when the time enters in the constraints and in the 
potential function, the dynamics of a mechanical system with an 
$n$ dimensional configuration space is related to a system of geodesics 
in an $(n+2)$ spacetime \cite{LPE,EM,GHKW}.
In order to simplify the problem, we shall assume that the constraits 
of the dynamical system and the potential $V$ do not 
involve time. In this simplified case it is adequately
to consider a Riemannian space with 
$n+1$ (in our particular case $4+1$) dimensions with the metric
\be\label{g5}
ds^2_5 = \sum^4_{i,j = 1} g_{ij}dq^i dq^j + A du^2\,,
\ee
where it is assumed that $A$ does not involve $u$.

In contrast with equations \eqref{geod}, now the trajectories of 
motion are given by
\ba\label{geodex}
&&g_{ij} \frac{d^2 q^i}{d s^2} + [jk,i] 
\frac{d q^j}{ds}\frac{d q^k}{ds}  -
\frac{1}{2}\frac{\partial A}{\partial q^i} \left( \frac{d u}{d s}
\right) ^2 = 0 
\,,\nonumber\\
&&A \frac{d u}{d s} = a \,,
\ea
where $a$ is a constant.

For a non vanishing constant $a$ it is possible to choose a parameter 
{\it t} for each non-minimal geodesic as
\be
t = a s\,,
\ee
identified with the time. Equations \eqref{geod} are the same as 
\eqref{geodex} if $A$ is defined by
\be\label{A}
\frac {1}{2 A} = V+ b\,,
\ee
where $b$ is another constant which should be chosen consistently with
\be
\frac{1}{a^2} = 2(E + b)\,.
\ee

At last, the coordinate $u$ is related to the action by
\be
u = -2 \int T dt + 2(E + b) t\,.
\ee

The Hamiltonian on the enlarge phase space \eqref{g5} is
\be
H_5 = \frac{1}{2}\sum_{i,j=1}^{4} g^{ij} p_i p_j + \frac{1}{2} 
\frac{1}{A} p_u^2 \,,
\ee
where $A$ is given by \eqref{A},  $p_i,p_u$ are the conjugate momenta 
and the new symplectic form is
\be
\omega' = dp_i \wedge d q^i  + dp_u \wedge d u \,.
\ee

Let us assume that the Hamiltonian \eqref{Hmusph} on 
$T^\star({\mathbb{R}}^4-\{0\})$ has a constant of motion polynomial in 
momenta of the form \eqref{cq}. We lift $K$ to the extended space
\be
{\cal K} = \sum_{i=0}^s p_u^{s-i} K^{(i)}\,.
\ee
It could be easily verified that $\cal{K}$ is a constant along 
geodesics on the enlarge phase space \eqref{g5} iff $K$ is a constant 
of motion for the original system \cite{GHKW}. In fact ${\cal K}$ is a 
homogeneous polynomial in momenta corresponding to a Killing tensor of 
the metric \eqref{g5}.

\section{Concluding remarks}

The aim of this paper is to use the covariant Hamiltonian formulation 
of the dynamics of particles in external gauge fields and scalar 
potential.

In general the explicit and hidden symmetries of a spacetime are 
encoded in the multitude of Killing vectors and higher order SK tensors 
respectively. The inclusion of gauge fields and scalar potentials 
affects the geodesic conserved quantities in a nontrivial way.

When we have a symplectic manifold with symmetries, it is possible to 
reduce the phase space to another symplectic manifold in which the 
symmetries are divided out. Such a situation arises when one has a 
particle moving in a gauge field $F$. If the group of symmetries acts on 
the manifold leaving the two-form $F$ invariant, it is possible to find 
a Hamiltonian system canonically induced on a reduced phase space.

In the usual applications, applying the method of reduction simplifies 
the equations of motion. However, in some cases, the 
reverse of the reduction might be useful, namely the equations of 
motion on the extended phase space are quite transparent, but the 
equations of motion of the quotient system appear more complicated. 
Applying an oxidation of a dynamical system with constants of motion 
polynomial in momenta, one may obtain spacetimes admitting SK tensors 
of higher rank.

The systems considered in this paper present hidden symmetries 
described by SK tensors of rank $2$. However there are several examples 
of integrable systems admitting integrals of motion of higher order in 
momenta. Recently it has been introduced \cite{VE} a new superintegrable 
Hamiltonian as a generalization of the Keplerian one with three terms 
preventing the particle crossing the principal planes. A generalization 
of the Runge-Lenz vector is found and also independent isolating 
integrals quartic in the momenta are identified. An investigation of 
the Kepler problem on $N-$dimensional Riemannian spaces of non constant 
curvature was done \cite{BEHRR} in order to obtain maximally 
superintegrable classical systems.

Another natural generalization of the Killing vectors is represented by 
totally antisymmetric Killing-Yano (KY) tensors. KY tensors generate 
supercharges in the dynamics of pseudo-classical spinning particles and 
non standard Dirac operators which commute with the standard one. Given 
a KY tensor one can construct a rank $2$ SK tensor as a symmetric 
product of KY tensors. It would be interesting to investigate relations 
between KY tensors and hidden symmetries in the context of Hamilton 
reduction and oxidation.

\subsection*{Acknowledgments}
Support through CNCSIS program IDEI-571/2008  is acknowledged.



\begin{thebibliography}{99}
%
\bibitem{AM}
R. Abraham and J. E. Marsden,
{\it Foundations of mechanics} (Benjamin/ Cummings, New York, 
N. Y., 1978).
%
\bibitem{LPE}
L. P. Eisenhart,
{\em Annals Math.} {\bf 30}, 591 (1928).
%
\bibitem{vH}
J. W. van Holten,
{\em Phys. Rev. D } {\bf 75}, 025027 (2007). 
%
\bibitem{JPN}
J.-P. Ngome,
{\em J. Math. Phys.} {\bf 50}, 122901 (2009).
%
\bibitem{MV1}
M. Visinescu,
{\em Mod. Phys. Lett. A} {\bf 25}, 341 (2010).
%
\bibitem{MW}
J. Marsden and A. Weinstein,
{\em Rep. Math. Phys.} {\bf 5}, 121 (1974).
%
\bibitem{MSS}
G. Marmo, E. J. Saletan and A. Simoni,
{\em J. Math. Phys.} {\bf 20}, 856 (1979).
%
\bibitem{KKS}
D. Kazdan, B. Kostant and S. Sternberg,
{\em Comm. Pure Appl. Math.} {\bf 31}, 481 (1978).
%
\bibitem{DV}
C. Duval and G. Valent,
{\em J. Math. Phys.} {\bf 46}, 053516 (2005). 
%
\bibitem{JMS}
J.-M.  Souriau, {\it Structures des Syst\`emes Dynamiques}
(Dunod, Paris, 1970).
%
\bibitem{HN}
P.A. Horv\'athy and J.-P. Ngome,
{\em Phys. Rev. D} {\bf 79}, 127701 (2009). 
%
\bibitem{IKI}
T. Igata, T. Koike and H. Isihara,
{\em Phys. Rev. D} {\bf 83}, 065027 (2011).
%
\bibitem{MV3}
M. Visinescu, 
{\em SIGMA} {\bf 7}, 037 (2011).
%
\bibitem{IU}
T. Iwai and Y. Uvano,
{\em J. Math. Phys.} {\bf 27}, 1523 (1986).
%
\bibitem{DZ}
D. Zwanziger,
{\em Phys. Rev.} {\bf 176}, 1480 (1968).
%
\bibitem{MIC}
H. V. McIntosch and A. Cisneros,
{\em J. Math. Phys.} {\bf 11}, 896 (1970).
%
\bibitem{SWH}
S. W. Hawking,
{\em Phys. Lett. A} {\bf 60}, 81 (1977).
%
\bibitem{NSM}
N, S, Manton,
{\em Phys. Lett. B} {\bf 110}, 54 (1982).
%
\bibitem{GR}
G. R. Gibbons and R. J. Ruback,
{\em Commun. Math. Phys.} {\bf 115}, 267 (1988).
%
\bibitem{IK1}
T. Iwai and N. Katayama,
{\em J. Math. Phys.} {\bf 36}, 1790 (1995).
%
\bibitem{IK2}
T. Iwai and N. Katayama,
{\em J. Geom. Phys.} {\bf 12}, 55 (1993).
%
\bibitem{GW}
G. W. Gibbons and C. M. Warnick,
{\em J. Geom. Phys.} {\bf 57}, 2286 (2007).
%
\bibitem{NH}
N. Hitchin,
{\it Monopoles, Minimal Surfaces and Algebraic Curves} 
(S\'emi\-nai\-re de Math\'ematiques Sup\'erieures {\bf 105}, 
Les Presses de L'Universit\'e de Montr\'eal, Montr\'eal, Canada, 1987).
%
\bibitem{MS}
M. Szydlowski,
{\em Gen. Rel. Grav.} {\bf 30}, 887 (1998).
%
\bibitem{IMB}
I. M. Benn,
{\em J. Math. Phys.} {\bf 47}, 022903 (2006).
%
\bibitem{EM}
E. Minguzzi, 
{\em Class. Quantum Grav.} {\bf 24}, 2781 (2007).
\bibitem{GHKW}
G. W. Gibbons, T. Houri, D. Kubiznak and C. M. Warnick,
{\em Phys. Lett. B} {\bf 700}, 68 (2011).
%
\bibitem{VE}
P. E. Verrier and N. W. Ewans,
{\em J. Math. Phys.} {\bf 49}, 022902 (2008).
%
\bibitem{BEHRR}
A. Ballesteros, A. Enciso, F. J. Herranz, O. Ragnisco and D. Riglioni,
{\em SIGMA} {\bf 7}, 048  (2011).
%
\end{thebibliography}
\end{document}